\documentstyle{mn}
\def\spose#1{\hbox to 0pt{#1\hss}}
\def\simlt{\mathrel{\spose{\lower 3pt\hbox{$\mathchar"218$}}
     \raise 2.0pt\hbox{$\mathchar"13C$}}}
\def\simgt{\mathrel{\spose{\lower 3pt\hbox{$\mathchar"218$}}
     \raise 2.0pt\hbox{$\mathchar"13E$}}}

\def\hmpc{\;h^{-1}{\rm Mpc}}
\def\h2mpc{\;h^{-2}{\rm Mpc}}
\def\hmpccc{\;h^{3}{\rm Mpc}^{-3}}

\def\kms{{\rm km\;s}^{-1}}

\def\bj{b_{\rm J}}

\def\degree{\nobreak\ifmmode{^\circ}\else{$^\circ$}\fi}

\begin{document}
\title[The Richness Dependence of Galaxy Cluster Correlations]
{The richness dependence
of galaxy cluster correlations: Results from a redshift survey of
rich APM clusters.}
\author[R.A.C. Croft {\it et al.}]{R.A.C. Croft$^{1,2}$
, G.B. Dalton$^1$, G.
Efstathiou$^1$, W.J. Sutherland$^1$\\
\vspace{-1mm}\\
{\LARGE and S.J. Maddox$^3$}\\
$^1$Department of Physics, University of Oxford,  
Keble Road, Oxford, OX1 3RH, UK.\\
$^2$Department of Astronomy, The Ohio State University,
Columbus, Ohio 43210, USA.\\
$^3$Royal Greenwich Observatory, Madingley Road, Cambridge,
CB3 0EZ, UK. \\
}
\maketitle
\begin{abstract}
We analyse the spatial clustering properties of a new
catalogue of very rich galaxy clusters selected from the APM Galaxy Survey.
These clusters are of comparable richness and space density
to  Abell Richness Class $\geq 1$ clusters,
 but selected using an objective algorithm
from a catalogue demonstrably free of artificial inhomogeneities. 
Evaluation  of the two-point correlation function $\xi_{cc}(r)$ for 
the full sample and for
richer subsamples reveals that the correlation amplitude is consistent with
that measured for lower richness APM clusters and X-ray selected clusters.
We apply a maxmimum likelihood
estimator to find the best
fitting slope and amplitude of a power law fit to  $\xi_{cc}(r)$, and
to estimate the correlation length
$r_{0}$ (the value of $r$ at which $\xi_{cc}(r)$ is equal
to unity). For clusters with a mean space density of 
$1.6\times 10^{-6}\hmpccc$ (equivalent to the space density of
Abell Richness $\geq 2$ clusters),
we find $r_{0}=21.3^{+11.1}_{-9.3} \hmpc$ ($95\%$ confidence limits). This is
consistent with the weak richness dependence of $\xi_{cc}(r)$ expected
in Gaussian models of structure formation. In particular,
the amplitude of  $\xi_{cc}(r)$
 at all richnesses matches that of $\xi_{cc}(r)$
for clusters selected in N-Body simulations of a low density Cold Dark 
Matter model. 
     
\end{abstract}
\section{Introduction}

Rich clusters of galaxies have been used by many authors as 
 tracers of the large-scale structure
of the Universe. Most analyses to date have relied on 
the cluster catalogue of Abell (1958) (and later Abell, Corwin \& Olowin 1989
 (ACO)).
Angular clustering statistics  for Abell clusters were calculated 
by Bogart \& Wagoner (1973) and Hauser \& Peebles (1973) and more recently,
 various redshift 
surveys of Abell clusters have been used to estimate the two point
cluster correlation function $\xi_{cc}(r)$ (eg. Bahcall \& Soneira 1983,
 Klypin \& Kopylov 1983,
 Postman, Huchra \& Geller 1992, Peacock \& West 1992).   
>From these studies, the two point correlation function for clusters has been
 found to be consistent in shape with the  power law form
measured for galaxies,
\begin{equation}
\xi_{cc}(r)=\left(\frac{r}{r_{0}}\right)^{-\gamma}.
\end{equation}
with a similar value of the power-law index $\gamma\sim 2$ but with a
higher amplitude $r_{0}$. For example, Peacock \& West (1992) find $r_{0}=21
\hmpc$ (where H$_{0}=100h \kms$) for Abell clusters of richness $R
\simgt 1$ whereas $r_{0}$ is around $5 \hmpc$ for galaxies (see eg.
Davis \& Peebles 1983).  Many authors have found, however, that there is much
evidence to suggest that the Abell catalogue, selected by eye from
unmatched photographic plates, is affected by inhomogeneities in
cluster selection which result in articifial clustering (Sutherland 1988;
Sutherland \& Efstathiou 1991; Dekel {\it et~al.} 1989; Peacock \& West 1992).

New results on the distribution of clusters have been obtained from an
automatically selected catalogue based on the APM Galaxy Survey
(Dalton {\it et~al.} 1992, hereafter DEMS92), 
and from smaller samples of clusters
selected from the Edinburgh--Durham Galaxy Catalogue and from the
ROSAT X-ray cluster survey (Nichol {\it et~al.} 1992; Romer {\it et~al.} 1994)
.  The amplitude of $\xi_{cc}$ measured from these studies is generally lower 
than for the Abell samples, so that $13 \hmpc \simlt r_{0} \simlt 16 \hmpc$.
However, it has been argued that the clustering seen in the automated
surveys is dominated by poor clusters, and that the results 
may be compatible with the higher values of $r_{0}$ measured for $R \simgt 1$
 Abell clusters, provided that
there is a strong dependence of the correlation length on cluster
richness. Bahcall \& West(1992) and Bahcall \& Cen (1992) 
argue that the Abell data are consistent with a linear
relation  between $r_{0}$ and mean intercluster separation $d_{c}$ 
($d_{c}=n_{c}^{-1/3}$ where $n_{c}$ is the mean space density) so that 
\begin{equation}
r_{0}=0.4 d_{c}.
\end{equation}
The evidence for this scaling relation, especially at high values of $d_{c}$
comes exclusively from estimates of the correlation functions of rich 
Abell clusters  (eg. Peacock \& West 1992). The validity of equation (2) thus 
depends critically on the uniformity of the Abell catalogue, particularly
at richnesses  $R \simgt 1$. The main aim of this paper is to test the scaling
relation (2) using an independent sample of rich clusters of galaxies selected
from the APM galaxy survey.

Croft \& Efstathiou (1994) have shown that the amplitude of the
cluster correlation function is is predicted to vary only weakly with
cluster space-density for a range of Cold Dark Matter (CDM) models. The
existing data for APM clusters (Dalton {\it et~al.} 1994a)
 are in good agreement 
with these predictions, but as the clusters are of relatively low
richness and hence low $d_{c}$ they are also consistent with the relationship
given in Equation 2. In  the study presented
here we use a new extension of the APM cluster survey  to test the behaviour
of $\xi_{cc}$ for richer clusters. 

The layout of this paper is as follows.
We describe the cluster sample and its
relationship to the samples of 
Dalton {\it et~al.}(1994a) and Dalton {\it et~al.} (1994b) in Section 2. In
Section 3 we present the correlation function for the new
cluster sample and for various subsamples
. We use a maximum likelihood estimator to fit a power law to the
correlation function and investigate how the fitted parameters change with
cluster richness. In Section 4 we compare our results with
other data samples and with N-body simulations of cosmological models.
We summarise our findings in
Section 5.

\section{The Cluster Sample}
DEMS presented a summary of the algorithm used to select galaxy
clusters from the APM survey. In Dalton {\it et~al.} (1994a) the selection
procedure was changed slightly in order to increase the volume
available to the survey.  A detailed description of the selection
procedure and a discussion of the effects of changing the various
selection parameters is presented in Dalton {\it et~al.} (1997).
 Here we use the
framework discussed in that paper to briefly describe changes
to the selection procedure which increase our sensitivity to rich
clusters at the expense of incompleteness at low richnesses.

The effective depths of the cluster samples selected from the APM
survey are limited by the definition of the cluster richness, ${\cal
R}$. This is defined to be the weighted number of galaxies in the
magnitude range $[m_X - 0.5,m_X + 1.0]$ above the mean background
count in the range $[m_X - 0.5,m_X + 1.5]$. Here $m_X$ is defined as
the magnitude of the galaxy for which the weighted count above
background exceeds $X={\cal R}/2$ for the catalogue of DEMS92 and
Dalton {\it et~al.} (1994b) .
  The depth of the cluster catalogue defined in this way
is fixed by the magnitude limit of the survey (b$_{J}$=20.5), so $m_X$
is constrained to be brighter than $b_{J}=19.0$. In Dalton {\it et~al.}
 (1994a) we
created a catalogue of greater depth by changing the background slice
to $[m_X -0.5,m_X+1.0]$, and redefining $X$ to be ${\cal R}/2.1$. By
combining this new catalogue with that of DEMS92 using a richness
transformation calibrated from clusters that appear in both catalogues
we created an extended sample of 364 clusters with APM richness ${\cal
R} \geq 50$ (sample B of Dalton {\it et~al.} 1994a). We will use results from
this comparatively low richness sample in comparisons with
measurements of clustering made from our new rich sample.

The new rich sample was created by further extending our survey by
increasing the limiting magnitude of the galaxy catalogue to
$\bj=21.0$.  The APM photometry is complete to this limit but the
fraction of objects which are mis-classified as stars rises sharply.
However the distribution of stellar objects fainter than b$_{J}$=20.5
is smooth on the scale of the counting annulus we use to determine the
backround correction, and so does not affect our ability to select
clusters.  We changed our cluster selection parameters to $X={\cal
R}/3$ and a richness counting slice (count and background) of
$[m_X-0.5,m_X+0.7]$ to optimise the depth increase gained by using the
extra $0.5\;{\rm mag}$ of galaxy data. We scaled the richness counts
in this catalogue by matching to the sample A catalogue in the same
way as for sample B (see Dalton {\it et~al.} 1994a) , and then targeted all
previously unidentified clusters with ${\cal R} \ge 80$.  In two and a
half clear nights at the Anglo-Australian Telescope (AAT) we were able
to obtain unambiguous redshifts for $100$ new clusters.  The redshifts
were obtained using the cross-correlation technique of
Tonry \& Davis (1979). The procedure is described in detail in Dalton 
{\it et~al.} (1994b).

To choose which clusters to observe, the catalogue was split into
different richness subsets and these were each split further into 4
bands by RA.  We were able to complete observations of the clusters in
all subsets except for the poor clusters with $RA > 2$ hrs. As the
efficiency of detection in our new catalogue is lowest for the poorest
clusters (${\cal R} \simlt 80$, due to the change in selection
parameters), most of the poor clusters in sample C are those
from the previous catalogue. Because of this, we limit our final
statistical sample to nearby clusters in order to have an essentially
complete sample. We choose to limit the sample to those clusters with
a redshift $cz \leq 55000 \kms$.  Above this redshift the overall
efficiency of cluster detection appears to fall rapidly as we
will show below.

Combining the cluster data with sample B gives a sample of $165$
clusters with ${\cal R} \ge 80$ which we shall refer to as sample
C. The redshift distribution of sample C is shown in panel (a) of
Fig. 1, together with smoothed distribution obtained by convolving the
histogram with a Gaussian of width $8000 \kms$.  The selection
function also shown (normalised to unity at the peak) suggests that
incompleteness in the deep sample starts to become important at high
redshift ($cz \ge 55000 \kms$).  The peak in the $n(z)$ distribution
at $cz=60000 \kms$ corresponds to a visible feature in the APM galaxy
map at $\alpha=23^{\rm h}, \delta=-20^\circ$, and appears to be
present in the ACO catalogue in the form of a large number of distance
class 6 clusters without published redshifts.  There is also a lack of
clusters in sample C at low redshifts ($cz \simlt 10000 \kms$). This
is mainly a consequence of clusters appearing too large on the plane
of the sky for selection using percolation Dalton {\it et~al.} (1997).
There also do
not appear to be any nearby very rich clusters in other surveys which
overlap with the APM such as the SSRS galaxy redshift survey
(da Costa {\it et~al.} 1994).
This is a very small fraction of the volume of the
rich cluster survey and in any case, we choose to limit the rest of
our analysis to the clusters with $cz > 10000 \kms$.  The redshift
distribution for sample B is also shown, for comparison, in panel (b)
of Fig 1. The smoothing of the distribution in this case was carried
out using $4000 \kms$ Gaussian because of the higher space density of
objects.

\begin{table}[b]
\centering
\caption[junk]{\label{tbdens} The space density of clusters}
\begin{tabular}{ccrrrr}
Sample & & $N_c$ & $cz_{min}$ & $cz_{max}$ & $n_c(\hmpccc)$ \\ \hline &&&&\\
B & ${\cal R} \ge 50$ & 362 & 5000 & 55000 & $3.4 \times 10^{-5}$ \\
B & ${\cal R} \ge 70$ & 114 & 5000 & 55000 & $9.0 \times 10^{-6}$ \\
C & ${\cal R} \ge 80$ & 163 & 10000 & 85000 & $5.4 \times 10^{-6}$ \\
C & ${\cal R} \ge 90$ &  79 & 10000 & 85000 & $3.1 \times 10^{-6}$ \\
C & ${\cal R} \ge 100$ &  37 & 10000 & 85000 & $1.8 \times 10^{-6}$ \\
C & ${\cal R} \ge 110$ &  18 & 10000 & 85000 & $1.6 \times 10^{-6}$ \\ \hline &&&&\\
\end{tabular}
\end{table}

We have estimated  the mean space density of clusters in this sample,
 using Equation 3 of Efstathiou {\it et~al.} (1992) and the results are
given in Table 1. We have also applied successively higher  richness 
bounds to create subsamples with lower space densities, the estimated space
densities as listed in  Table 1.
We also list the space density of sample B and a subsample  with  
${\cal R} \ge 70$.

\begin{figure}
\centering
\vspace{8.0cm}
\includegraphics{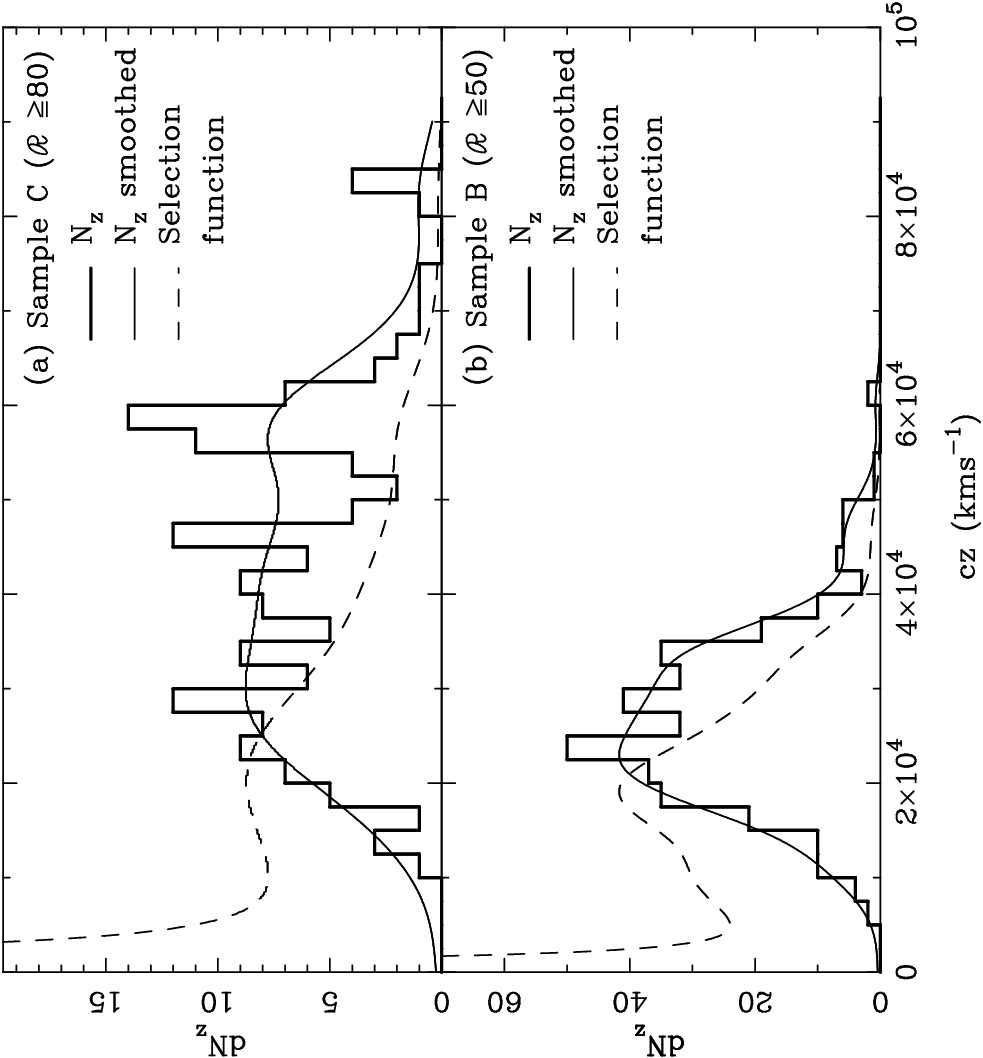}
\caption[junk]{\label{junk3} (a) The redshift distribution of Sample C,
the new rich cluster sample. The thick lines are a histogram of the
distribution of cluster redshifts, a smoothed version of which (see text)
is shown by the thin solid line. The dashed line represents the selection
function for the sample, obtained by dviding the smoothed distribution by
the appropriate volume element.(b) For comparison, we plot dN$_{z}$ for the
survey of the 
extended sample of 364 clusters (Sample B) of Dalton {\it et~al.} (1994a)}
\end{figure}

\section{Cluster Correlations}
We estimate the redshift-space correlation functions for the
samples in Table 1 by cross-correlating with a random catalogue and
using the estimator
\begin{equation}
\label{eqxidef}
\xi_{cc}(s) = 2f\frac{DD}{DR} -1, 
\end{equation}
where DD and DR are  the number of cluster-cluster pairs and the number of 
cluster-random pairs respectively in each bin centred on $s$. 
The parameter $f$ is the ratio of the number of random points to the number of
clusters in the sample. In each case we use $20,000$ points
distributed within the survey boundaries and with the same redshift
distributions as the smoothed distributions shown in
Figure 1. As stated in section (2), we  present results for the 
clusters with  $cz \le 55000 \kms$ in order to be minimise any effects that
are due to uncertainty in the selection function
at high redshift. This being the case, we have also studied  
the clustering of the full sample and in all
cases find the measurements to lie within $1\sigma$ of the
 $cz \le 55000 \kms$ sample results.

We  also use the estimator of :Hamilton (1993):
\begin{equation}
\label{eqxidef}
\xi_{cc}(s) = 4\frac{(DD)(RR)}{(DR)^{2}} -1, 
\end{equation}
which is less affected by uncertainties in the selection function for
$\xi_{cc}<1$.

\begin{figure}
\centering
\vspace{8.7cm}
\includegraphics{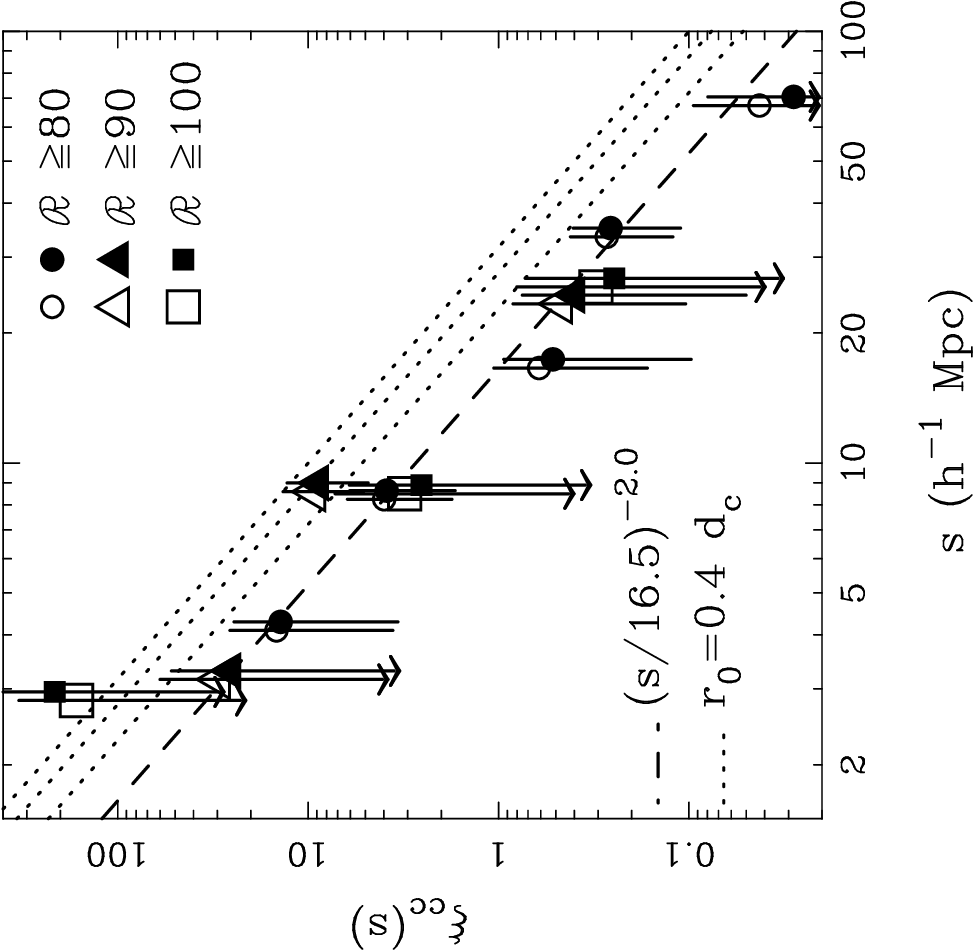}
\caption[junk]{\label{junk4} 
The two-point correlation function for the three sub-samples of clusters
from  sample C, as discussed in the text. 
The estimator of equation (4) was used to calculate the solid symbols and 
equation (3) for the open symbols (which have been displaced to the left
slightly to make the error bars visible).
The dashed line 
represents the best fit to the data for  ${\cal R} \ge 80$,
$\xi_{cc}=(s/16.5 \hmpc)^{-2.0}$. 
The dotted lines show the prediction of equation (2) for the power-law
fits to the correlation function of each of the sub-samples.
}
\end{figure}

To test the dependence of the correlation function on cluster space
density we estimate $\xi_{cc}(r)$ for subsamples with ${\cal R} \ge
90$ and ${\cal R} \ge 100$. In Figure 2 we show the correlation
functions for the full sample and for these subsamples.  We note that
for the ${\cal R} \ge 100$ clusters each point is within $1\sigma$ of
zero.  A least-squares fit to the data (using estimator (3)) for
${\cal R}\ge 80$ yields $\gamma=2.0\pm0.4$ and $A=10^{2.4\pm0.44}$
where $A=r_{0}^{\gamma}$, and the quoted errors are $1\sigma$. This
gives $r_{0}=16.5\hmpc$. If the slope is constrained to be
$\gamma=2.0$ we find $r_{0}=16.5^{+7.0}_{-6.0}\hmpc$ where the errors
are calculated from the 5 percentile points of the $\chi^{2}$
distribution.  The best fit power law for the ${\cal R} \ge 90$
subsample is steeper, with $\gamma=2.8\pm1.0$, $A=10^{3.8\pm1.0}$. If
$\gamma$ for this subsample is constrained to be $2.8$, then we obtain
$r_{0}=19.1^{+8.0}_{-7.5}\hmpc$.

The fit for ${\cal R} \ge 80$ is shown as the dashed line in Figure
2. The data for the ${\cal R} \ge 100$ and ${\cal R} \ge 90$ sample
appear to be in reasonable agreement with the ${\cal R} \ge 80$
sample. 
We have plotted the predictions of
equation (2) for the correlation functions of the different samples.
Most of the points lie below the corresponding prediction.  We
therefore conclude that the correlation amplitude is not strongly
dependent on the cluster richness.

\subsection{A maximum likelihood estimator of $\gamma$ and $r_{0}$.}

As we are interested in the behaviour of $r_{0}$ as a function of cluster 
richness, we would like to be able to estimate its value and error bounds
in the most direct way possible. Binning the data introduces uncertainties,
as the value of $\xi(r)$ can depend on the binning interval and the position
of bin centres (in log or linear space). We circumvent these
problems by maximising the likelihood that a power law form for $\xi(r)$, 
as in equation (1)  will produce the observed set of pair
separations. In this way, we can find confidence limits on the 
two parameters $\gamma$ and $r_{0}$
, even for small numbers of clusters.

To construct our estimator, we need to find the predicted probability 
distribution of cluster pairs for each value of $\gamma$ and $r_{0}$.
We  deal with the mask and selection function in the usual way by
creating a catalogue of random points 
 with the same boundaries and selection
function as the cluster catalogue in question. We then calculate the 
separations of all the cluster-random pairs and bin them in $r$.
The bin width can be made arbitrarily small as long as number of
points in the random catalogue is increased accordingly. In this case
we use 100000 random points in the catalogue
and 200 bins in the interval $0-100 \hmpc$. 
If the 
number of cluster-random pairs in an interval $dr$ is $g(r)dr$,
then the predicted mean number of cluster-cluster pairs in that interval
is $h(r) dr$ where
\begin{equation}
h(r) dr = f (1+\xi_{cc}(r))g(r) dr, 
\end{equation}
$f$ is the number of clusters divided by the number of random points, and
$\xi_{cc}(r)$ has the power law form given by Equation 1. 
We can then use the separations ($r_{i}$) of all the ($N$) cluster-cluster
pairs to form a likelihood function ${\cal{L}}$.
${\cal{L}}$ is defined as the product of the probabilities of having
exactly one pair in the interval $dr$ at each of the pair separations $r_{i}$
of the N pairs and the probabilty of having zero pairs in all the other
differential elements of r. This is for all r in a chosen range
(say $r_{a}$ to $r_{b}$), in
our case the range of values for which $\xi_{cc}(r)$ can be reasonably
expected to have power law behaviour.
To find the likelihood, we assume Poisson probabilties, so that
(see also Marshall {\it et~al.} 1983):

\begin{equation}
{\cal{L}}=\prod_{i}^{N}e^{-\mu}\mu \prod_{j\ne i}e^{-\mu}, 
\end{equation} 
where $\mu=h(r) dr$, the expected number of pairs in the interval $dr$, and
the index $j$ runs over all the elements $dr$ in which there are no pairs.  
We then define the usual quantity $S=-2\ln {\cal{L}}$  and
drop all terms independent of model parameters, so that 

\begin{equation}
S=2\int^{r_{b}}_{r_{a}} h(r) dr - 2\sum^{N}_{i}\ln(h(r_{i})).
\end{equation}
 
The best fit values of $r_{0}$ and $\gamma$ are obtained by minimising 
$S$, with confidence levels defined by 
 $\Delta_{S}=S(r_{best},\gamma_{best})-S(r_{0},\gamma)$, assuming
that $\Delta_{S}$ is distributed with a $\chi^{2}$ distribution
. These confidence limits are likely to be 
underestimates, as the  assumption of Poisson statistics 
assumes that
all pairs are independent of each other. It may be possible to incorporate
the effects of higher order correlations into the likelihood by using
a scaling model for the three point and higher correlation functions 
(see e.g. Peebles 1980). This would 
result in more accurate error bars. 
However, we can use N-body simulations to give us an idea of the real 
errors.
 Croft \& Efstathiou (1994) compared  error
bars on the individual points obtained using Poisson statistics
with the scatter between results for different simulated 
cluster surveys. In that case, the ensemble errors, which 
include the additional effects of cosmic variance were between 1.3 and 1.7
times larger than the Poisson errors. We expect the errors computed 
from our likelihood analysis to be underestimates by roughly the same 
factor. We  check here whether this is the case by using 
large box size N-body simulations (see Section 4) to make simulated 
cluster catalogues 
with the same angular shape and  selection function as sample C and
richer subsamples selected from it (see Section 3). We have done this
for the low density CDM model (see Section 4) for which $10$ 
simulations are 
available. In Table 2 we present the values of $r_{0}$ and $\gamma$ and
their $1\sigma$ confidence intervals (($\delta_{r0})_{l}$ and
$(\delta_{\gamma})_{l}$)  obtained by applying the maximum
 likelihood estimator to the simulated
catalogues. Also shown is the ratio of these errors to the ensemble errors
($\sigma(r_{0})$ and $\sigma(\gamma)$). From these results we can see that
our expectations are approximately correct and that the likelihood errors
are underestimates by between $1.1$ and $2.1$. We can also see that 
the Poisson errors are closer to the real errors when the number of clusters
is small. Our estimates of the error bars for the richest subsamples of
clusters should therefore be the most accurate.

\begin{table}[b]
\centering
\caption[ta3]{ Errors estimated from the likelihood 
distribution  compared to  the scatter between 10 simulations of a
low density CDM universe. The errors on
$\gamma$ and $r_{0}$ are the average of the $1\sigma$ confidence
intervals ( $(\delta_{r0})_{l}$ 
and $(\delta_{\gamma})_{l}$) obtained by
applying the maximum likelihood method to 10 simulated catalogues.
The standard deviations of the measurements taken from the simulated catalogues
are denoted by $\sigma_{r0}$ and $\sigma_{\gamma}$.  
\label{ta3}}
\vspace{0.2cm}
\begin{tabular}{ccccc} 
$\hmpc$ &   $\hmpc$       &   &                   &             \\
$d_{c}$ &$r_{0}\pm 1\sigma$ &$(\delta_{r0})_{l}/\sigma(r_{0})$   &$\gamma \pm  1\sigma$ & $(\delta_{\gamma})_{l}/\sigma(\gamma)$            \\ \hline &&&&\\
   57   & $17.2^{+1.5}_{-1.6}$& 0.48 & $1.97^{+0.20}_{-0.20}$& 0.76\\
   69   & $18.1^{+2.8}_{-3.4}$& 0.60 & $1.93^{+0.36}_{-0.36}$& 0.90\\
   79   & $18.2^{+5.2}_{-6.7}$& 0.88 & $1.94^{+0.70}_{-0.73}$& 0.93\\ \hline
\end{tabular}
\end{table}

\subsection{$\gamma$ and $r_{0}$ from APM clusters and the richness dependence
of $r_{0}$.}

\begin{figure*}
\centering
\vspace{13.0cm}
\includegraphics{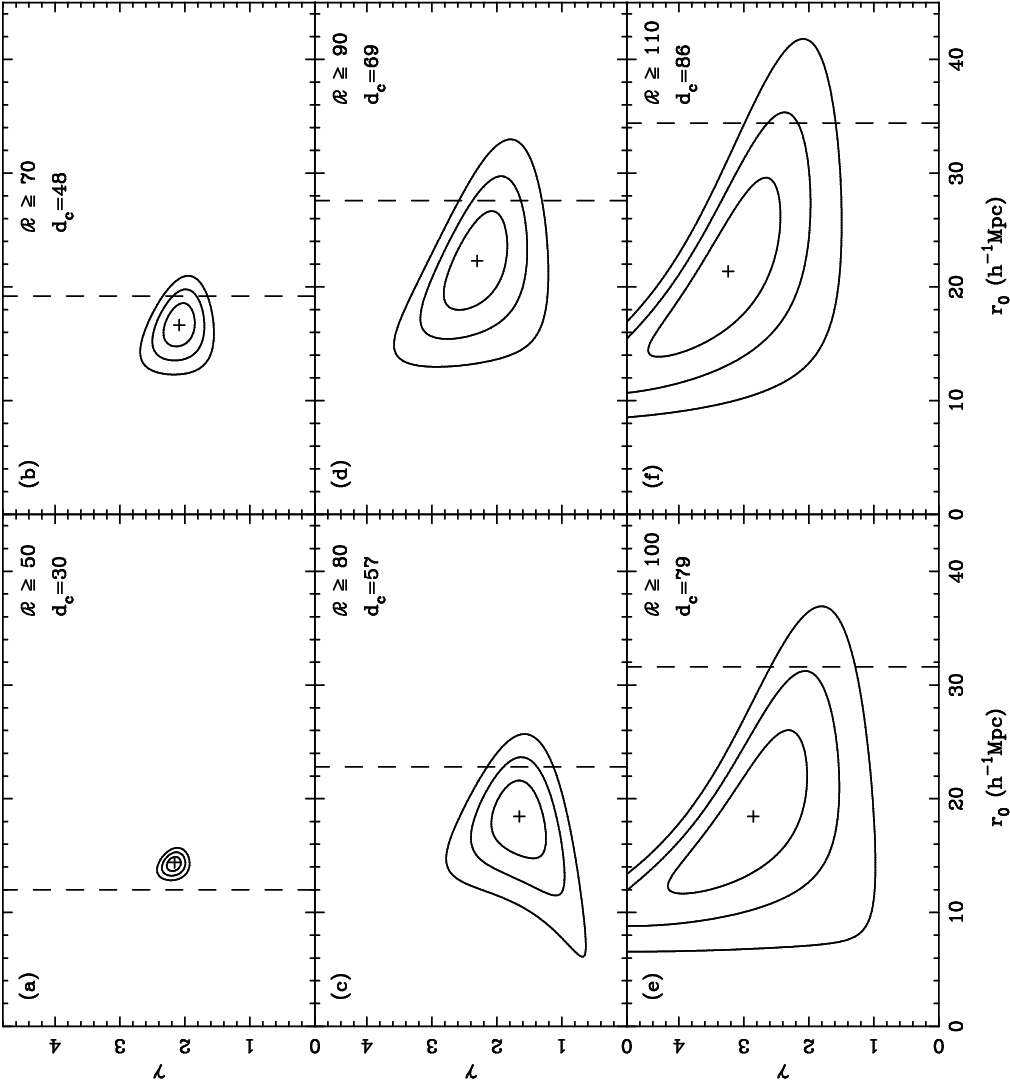}
\caption[junk]{\label{junk2} 
Contours  resulting from maximum likelihood analysis
of cluster pair separations in order to find the most probable values of
$r_{0}$ and $\gamma$. The best fit values of these two parameters
are shown by a cross in each plot.
The contours enclose $68 \%$, $95\%$ and $99.7\%$ 
of the joint probability respectively, if the distribution of 
$S=-2\ln {\cal{L}}$ follows a $\chi^{2}$ distribution with 2
degrees of freedom.
The different panels show results for different subsamples of
clusters, with varying lower richness limits.
Panel (a) shows results for the extended sample of 364 clusters 
of Dalton {\it et~al.} (1994a). Panel (b) shows results for the
clusters in sample B with ${\cal{R}}\geq 80$.
 In panels (c) to (f) we use the new rich sample (C)
 that forms
the subject of this paper. In each case the lower richness limit and the
resulting mean intercluster separation $d_{c}$ are shown in the top
right of the panel. 
incompleteness. The dashed lines show the relation r$_{0}= 
0.4 d_{c}$ of Bahcall \& West (1992).}
\end{figure*}

We apply the estimator described above to the catalogue C subsample of
110 rich APM clusters with $cz < 55000 \kms$ as well as subsamples
with varying lower richness bounds. From consideration of the plot of
$\xi_{cc}(r)$ in bins for the full sample, as well as the results for
sample B of Dalton {\it et~al.} (1994a) , we decide that $\xi_{cc}(r)$ can
probably be fitted to a power law over the range $r_{a}=2\hmpc$ to
$r_{b}=70\hmpc$. Varying these limits does not greatly affect the
results, although if $r_{b}$ is increased to much over $70-80 \hmpc$,
the fitted power law steepens slightly, probably due to a break in
$\xi_{cc}(r)$.

We also apply the estimator to sample B, the results for which are
shown in the first panel of Figure 3. In this case, we find
$r_{0}=14.2^{+0.8}_{-1.0} \hmpc$ and $\gamma=2.13^{+0.16}_{-0.14}$
where the errors indicate the $95\%$ confidence bounds on each
parameter individually.  The contours on the plot show the joint
confidence bounds at levels of $68 \%$, $95 \%$ and $99.7 \%$.  If we
choose to constrain $\gamma=2.13$ and find the maxmimum likelihood
value of $r_{0}$ in one dimension we also get
$r_{0}=14.2^{+0.8}_{-1.0} \hmpc$ at $95 \%$ confidence.  The
$\chi^{2}$ fits to the binned $\xi(r)$ give
$\gamma=2.05^{+0.20}_{-0.20}$ ($2\sigma$ errors).  If the slope is
constrained to have this value, then from the binned data
$r_{0}=14.3^{+2.5}_{-2.25} \hmpc$ (Dalton {\it et~al.} 1994a).  The errors
on $r_0$ obtained from the binned data are therefore a factor of 2
larger than the errors from the maxmimum likelihood technique.

We have investigated a few possible reasons for this discrepancy.
%
%
%
The main reason appears to be
an anomalously low $\chi^{2}$ for the power law fit.
Fitting to the 7 bins above $2 \hmpc$ we find $\chi^{2}=1.7$,
which should only occur $\sim 10\%$ of the time. The binned data
for the richer subsamples have more normal values of $\chi^{2}$ and
errors much closer to the maximum likelihood values.
In discussing our results we will 
concentrate our attention on the fit parameters derived using the maximum
likelihood method.

We have applied the maximum likelihood method to a subsample of clusters
from sample B with ${\cal R} \ge 70$, with the results shown in  Figure 3(b).
 These clusters have a mean separation 
$d_{c}=48 \hmpc$ and have a slightly larger value of $r_{0}=16.6\pm2.6\hmpc$.
We tabulate the best fit 
parameters $\gamma$ and $r_{0}$ for this and all the other
cluster samples in Table 3. We also present the $1 \sigma$ and $2 \sigma$
confidence limits on each parameter taken individually.

The results for sample C and subsamples of higher richness are
shown in Figure 3(c)-(f).  These subsamples are the same as those used in 
calculating the binned  correlation functions plotted in Figure 2, with the
addition of a subsample of APM clusters with ${\cal R} \ge 110$.
The  APM ${\cal R} \ge 110$ clusters have a similar space density to Abell
$R\ge 2$ clusters (see e.g. Peacock \& West 1992).
We also plot a  dashed line showing
the relation $r_{0}=0.4d_{c}$ of Bahcall \& West (1992). 

\begin{table*}[t]
\centering
\caption[ta2]{$r_{0}$ vs. $d_{c}$ for different samples of APM clusters. 
\label{ta2}}
\vspace{0.2cm}
\begin{tabular}{cccccccc} 
       &   & Number of &  $\hmpc$   &$\hmpc$    &$\hmpc$&&           \\
Sample&  & clusters  &  $d_{c}$   &  $r_{0}\pm 1\sigma$  & $r_{0}\pm 
2\sigma$ & $\gamma \pm  1\sigma$ & $\gamma \pm  2\sigma$ \\ \hline
&&&&&&\\
B& ${\cal{R}}\geq 50$& 364&   30       & $14.2^{+0.4}_{-0.6}$&
 $14.2^{+0.8}_{-1.0}$& $2.13^{+0.09}_{-0.06}$&$2.13^{+0.16}_{-0.14}$\\
B& ${\cal{R}}\geq 70$& 114&   48       & $16.6^{+1.3}_{-1.3}$&
 $16.6^{+2.6}_{-2.6}$& $2.1^{+0.2}_{-0.2}$&$2.1^{+0.3}_{-0.3}$\\
C& ${\cal{R}}\geq 80$& 110&   57       & $18.4^{+2.2}_{-2.4}$&
$18.4^{+4.2}_{-5.1}$&$1.7^{+0.3}_{-0.3}$&$1.7^{+0.6}_{-0.6}$\\
C& ${\cal{R}}\geq 90$& 58&   69       & $22.2^{+2.8}_{-2.8}$&
$22.2^{+6.0}_{-5.5}$& $2.3^{+0.3}_{-0.3}$&$2.3^{+0.7}_{-0.7}$\\
C& ${\cal{R}}\geq 100$& 29&   79       & $18.4^{+4.8}_{-4.8}$&
$18.4^{+10.2}_{-8.4}$& $2.8^{+0.8}_{-0.6}$&$2.8^{+1.8}_{-1.1}$\\
C& ${\cal{R}}\geq 110$& 17&   86       & $21.3^{+5.3}_{-5.3}$&
$21.3^{+11.1}_{-9.3}$& $3.2^{+0.8}_{-0.6}$&$3.2^{+1.6}_{-1.1}$\\ \hline
\end{tabular}
\end{table*}

We can see from the contour plots that there is a slight anticorrelation
of $r_{0}$ and $\gamma$, so that lower values of $r_{0}$ would result
in a steeper slope for $\xi_{cc}(r)$. As the errors are large,
the value of $\gamma=2.1$ obtained for sample B with 364 clusters
is broadly compatible with $\gamma$ for the rich sample, C. The slope
of $\xi_{cc}(r)$ seems to be steeper than that normally
quoted for the correlation function of galaxies ($\gamma \simeq 1.8$ see
eg. Davis \& Peebles 1983).

\section{Discussion}

\subsection{Comparison with results for other cluster catalogues.}

A plot of $r_{0}$ versus $d_{c}$ for various observational samples
of clusters including the APM samples (taken from Table 2 above and
 DEMS92 and labelled APM and APM92 respectively).
is shown in Fig 4.
 For the maximum likelihood points, the errors are $1 \sigma $
for marginalisation of $r_{0}$ over all values of $\gamma$.
The APM92 points are for $\gamma$ constrained to be 2.0.
The points labelled `Abell' indicate the results for Abell $R \geq 0$
, Abell $R \geq 1$ and Abell $R \geq2$ clusters derived by 
Peacock \& West (1992). 
The point labelled EDCC is the result for 79 clusters from the
Edinburgh-Durham Cluster Catalogue of Lumsden {\it et~al.} (1992) estimated by
Nichol {\it et~al.} (1992). The error bar
size was estimated using bootstrap resamplings. The same is true of the 
error on the point labelled X-Abell, which was estimated by
Nichol, Briel \& Henry (1994),
from  67 clusters in
the redshift sample of Huchra {\it et~al.} (1990) which also have X-ray 
luminosities $\geq 10^{43}$ erg s$^{-1}$. 
The point labelled ROSAT shows $r_{0}$
for $\xi_{cc}(r)$ measured from a redshift survey of an X-ray flux 
limited sample of clusters (Romer {\it et~al.} 1994).
 The X-ray flux for both these
last samples was measured using the ROSAT satellite.

\begin{figure}
\centering
\vspace{7.5cm}
\includegraphics{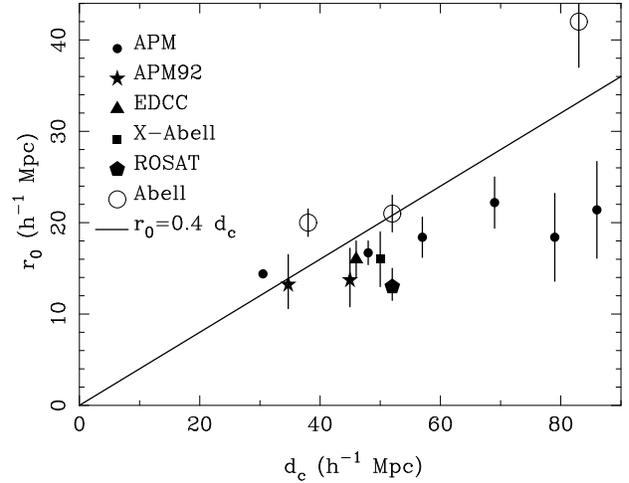}
\caption[junk]{\label{junk2} 
 The quantity $r_{0}$  (the correlation
length) plotted against cluster space density for a number of observed
cluster samples (see text).
Error bars represent the 1 $\sigma$ error on the mean.
The solid line shows the relation r$_{0}= 0.4 d_{c}$ of 
Bahcall \& West (1992).}
\end{figure}

It can be seen that most of the data points are for cluster samples
with $d_{c}$ in the range $30-55\hmpc$, and that in this range, the
results for the X-ray samples and automated galaxy surveys are in
agreement with one another, and lower than those for Abell
clusters. As has been detailed previously, this can be understood as
being due to non-uniformities in the Abell catalogue which artificially
boosts the amplitude of clustering. Over this small range in cluster
space density, for which the errors are comparatively small, there is
not much evidence for any trend of $r_{0}$ with $d_{c}$ and hence
cluster richness. Part of the reason for the work in this paper was to
find out whether this is also true at higher richnesses and lower
space densities. The solid line in the plot corresponds to the scaling
relation $r_{0}=0.4d_{c}$ proposed by Bahcall \& West (1992) as a fit to the
correlation functions of the Abell sample. As can be seen from the
plot, the motivation for assuming this fit at high values of $d_{c}$
was provided by the results for Abell $R \geq 2$ clusters (a sample of
42 clusters was used to calculate this data point -- see
Peacock \& West 1992). 

Now that we have a sample of very rich clusters taken from a catalogue
which is demonstrably free of artificial inhomogeneities, we are in
the position to test equation (2) using the APM data alone. The three
APM points on the right of the plot are for ${\cal{R}}\geq 90,
{\cal{R}}\geq 100 $ and ${\cal{R}}\geq 110$ clusters, which have space
densities comparable to that of the Abell $R \geq 2$ clusters. If the
error bars are taken at face value, then the relation would appear to
be ruled out at the $\sim 2 \sigma$ level. However, as we have seen
from Table 1, the error bars could be underestimates by a factor of
$\sim1.1 - 2.1$. Also, the space densities of clusters used to derive
$d_{c}$ values are not precise estimates because of the difficulties
involved in estimating the completeness of richness limited cluster
catalogues (see Efstathiou {\it et~al.} 1992).
 That said, we believe that these data
points are more reliable than those for the Abell $R \geq 2$
clusters. Table 2.1 also shows us that the error bars for the richest
sub-samples are likely to be the most accurate. In summary, the APM points are
consistent with a weak dependence of clustering on richness.
We find no evidence that equation (2) applies to rich clusters of
galaxies, with important implications for theories of structure
formation as described in the next section.

\subsection{Comparison with model predictions.}

Croft \& Efstathiou (1994)
examined the behaviour of $r_{0}$ with $d_{c}$ expected in
several popular cosmological scenarios (see also Bahcall \& Cen 1992,
 Mann, Heavens \& Peacock 1993).
 The 
box size ($300 \hmpc$) of the dissipationless
N-body simulations used in that study,
 meant that the predictions did not extend to the large values
 of $d_{c}$ needed to make comparisons with our new rich cluster sample. 
We have  therefore
run  a set of simulations (using the same particle-particle particle-mesh
N-body code) with  box size 
 $600 \hmpc$ and $4 \times 10^{6}$ particles.
These simulations are the same as those used in Croft \& Efstathiou (1995). 
The models we shall consider are the Standard CDM model 
(SCDM has $\Gamma= \Omega h=0.5$ and $\Omega=1$) and the spatially flat
Low density CDM model (LCDM has $\Gamma=0.2$, $\Omega=0.2$ and
$\Omega_{\Lambda}=0.8$). Both models are normalised to be compatible with
the first year 
COBE anisotropies (Wright {\it et~al.} 1994) so that $\sigma_{8}=1.0$ for both
models, where $\sigma_{8}$ is the rms amplitude of linear fluctuations 
in $8\hmpc$ spheres. The results are insenstive to the precise value
of $\sigma_{8}$ .
We use the same techniques as in Croft \& Efstathiou (1994) and
 Dalton {\it et~al.} (1994a)
 to find clusters in the 
simulations. This involves
finding cluster centres in real space with a percolation algorithm
and then ordering clusters by the mass contained within a 
certain  radius, in this case  $0.5 \hmpc$. We then calculate
$r_{0}$ for clusters with different lower mass limits, with the results 
shown in Figure 5. 
We have chosen to calculate the correlation functions in redshift
space, for more accurate comparison with the observations.
The values of $r_{0}$ which we present below
are $\sim 1 \hmpc$ larger than the values estimated in real space.

The correlation functions for the LCDM model are shown in Figure 5, together 
with the APM  points (estimated using Equation 4). The space densities
of the simulated clusters were selected to be close to those for the 
three subsamples of rich APM clusters plotted.
The curves plotted are the averages of results for 10 simulations of LCDM.
 We can see that the APM results are compatible with LCDM model. We can
 also see that the clustering strength of LCDM clusters increases only 
a small amount as the richness bound is increased.

\begin{figure}
\centering
\vspace{8.7cm}
\includegraphics{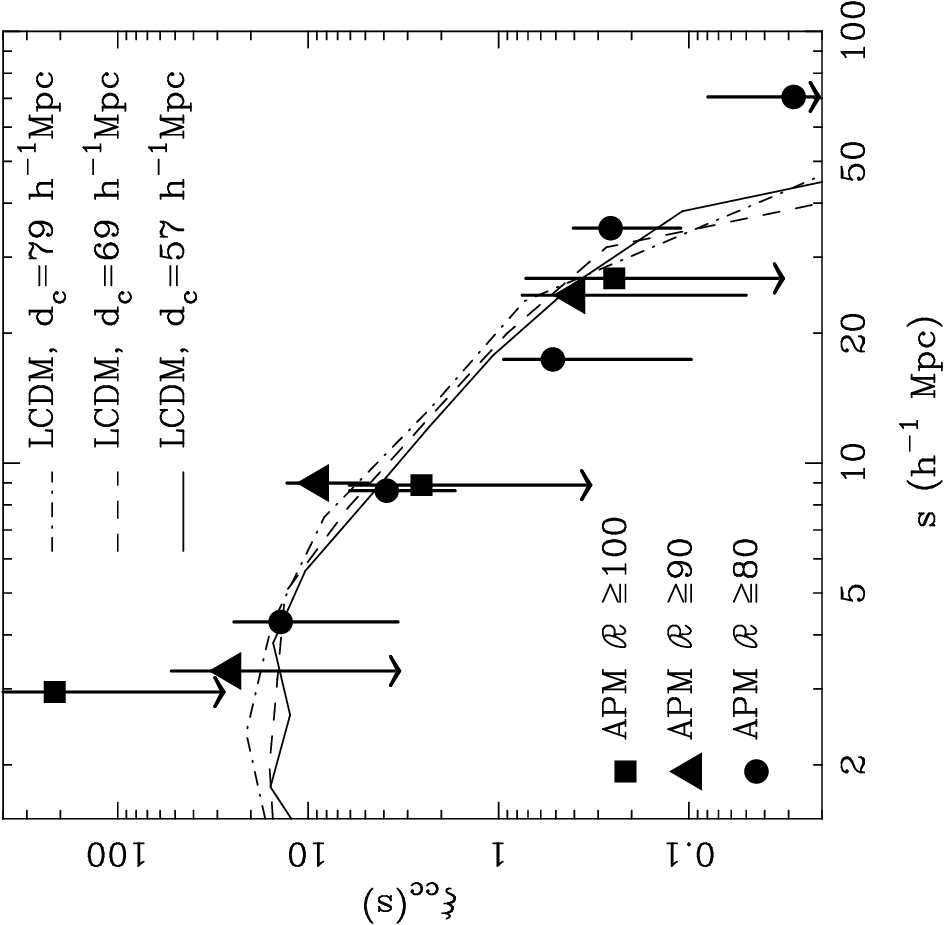}
\caption[junk]{\label{junk2}
The two-point correlation function of simulated clusters in the LCDM model
in redshift space for
subsamples with three different mean separations. The APM
results from Figure 2 (solid symbols, computed using the estimator of
Equation 4) are also shown.
}

\end{figure}

In order to see how the clustering results are affected by the mask and 
selection function, we have plotted the results for the mock APM cluster
catalogues constructed from 10 LCDM simulations and described
in Section 3.1. The results are shown in Figure 6 in the form of a scatter
plot. In each panel we plot $r_{0}$ against $\gamma$
(both measured from the maximum likelihood technique). We show results
calculated from clusters with the same $d_{c}$ values as those in Figure 5.
We also plot points for the APM results for equivalent richess clusters.
In each of the panels we can see that the APM results are not extreme outliers
and it looks plausible that they could have been drawn from the
same distribution as the LCDM points. A line denoting the relationship of
Equation 2 is drawn on each panel. From this we can conclude that in an
LCDM Universe we would have a $\sim 10 \%$ chance 
for each richness cut of measuring a value of $r_{0}$  which fits
this relationship.

\begin{figure}
\centering
\vspace{14.7cm}
\includegraphics{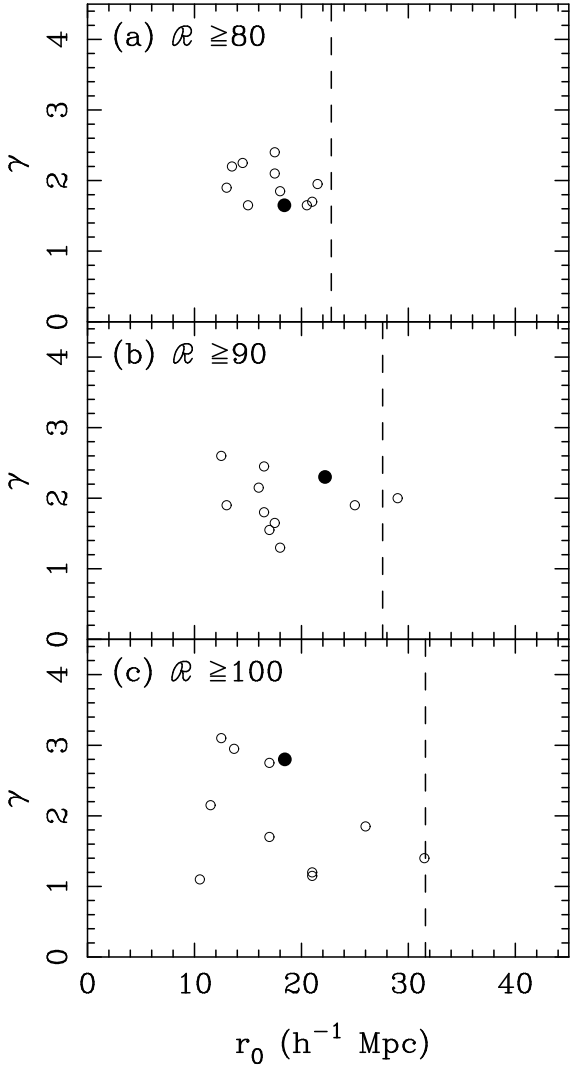}
\caption[junk]{\label{junk2}
Values of $r_{0}$ and $\gamma$ measured from mock catalogues
constructed from 10 simulations of an LCDM universe (see text).
Plotted are results for clusters
 with three different mean separations. The corresponding APM
results from Figure 3. are also shown.
}
\end{figure}

\begin{figure}
\centering
\vspace{7.5cm}
\includegraphics{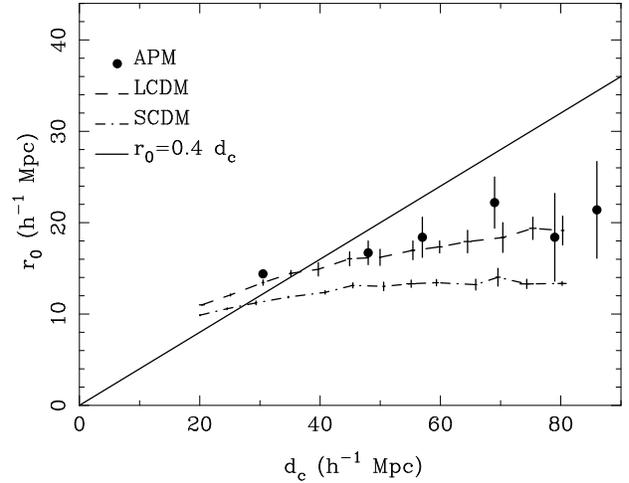}
\caption[junk]{\label{junk2} A comparison of the richness dependence
of APM cluster correlations (filled circles) with the corresponding
predictions for a low density CDM Model (dashed line) and Standard CDM
(dot-dashed) line. The theoretical predictions have been calculated in
redshift space.
Error bars represent the 1 $\sigma$ error on the mean.
The solid line shows the relation r$_{0}= 0.4 d_{c}$ of 
Bahcall \& West (1992).}
\end{figure}

In Figure 7 we plot $r_{0}$ measured from  the correlation functions
of the LCDM and SCDM clusters against $d_{c}$. 
 The error bars on the simulation points were calculated from the $1 \sigma$ 
error on the mean taken from 3 simulations of each model. We therefore
have 2.4 times as many clusters of any given space density as in the ensembles
of Croft \& Efstathiou (1994).
We also plot the values of $r_{0}$ calculated using the maximum likelihood
method in this paper.
 The plots shows the very weak trend of clustering strength with cluster
richness continuing for both models at least up to $r=80 \hmpc$.  The APM
points are 
consistent with the LCDM model, but not with SCDM. We note here that a
simulation with a
different amplitude of clustering in the underlying mass could have 
an $r_{0}$ which differs by as much as $1-3 \hmpc$
 as could clusters which are selected using a different method.
These variations are not expected to be large enough to
affect our conclusions (Croft \& Efstathiou 1994,
Eke {\it et~al.} 1996, Mo, Jing \& White 1996).

It is encouraging that models with $\Gamma \approx 0.2$, which were introduced
to explain clustering in the galaxy distribution
(see eg. Efstathiou, Sutherland \& Maddox 1990)  are also able
to match well  the clustering of rare and extreme objects 
such as the rich galaxy clusters considered here. 
We also expect other Gaussian models with similar power spectra
such as a Mixed Dark Matter (MDM) universe dominated by CDM and with
an additional component of massive neutrinos (see eg. Klypin {\it et~al.} 
1993) 
to be compatible with our APM results at high richnesses, as
they are at low richnesses (Dalton {\it et~al.} 1994a). 
>From Figure 7 we can also see that whilst models such as low density CDM
provide a good fit to the clustering behaviour of rich APM clusters
they are completely incompatible with the scaling relation derived from 
considering rich Abell clusters. 
Our data exclude such a strong scaling relation and remove the
need to resort to non-Gaussian models for the formation of large-scale 
structure.

\section{Summary}
We have carried out a new  extension of
the APM cluster redshift survey to provide a
sample of 165 clusters with richnesses ${\cal{R}}\geq 80$ and
mean space density of $5.4\times 10^{-6} \hmpc^{-3}$. The correlation function
of this sample is found to be consistent with the clustering 
amplitude measured for our previous larger sample of poorer APM
clusters. Restricting the evaluation of $\xi_{cc}(r)$ to even 
 richer subsamples shows that there is 
only a weak dependence of correlation length 
with cluster richness. This is disagrees with 
results from Abell $R\geq2$ clusters. 
The high amplitude of $\xi_{cc}$ for the Abell $R \ge2$ sample
is most probably caused by inhomgeneities in the Abell catalogue. The
weak dependence of clustering strength with richness that we find in the
APM survey is however in good agreement with what is
 expected in a universe with Gaussian
initial fluctuations and a power spectrum with more large-scale power
than standard CDM, such as low density CDM or MDM.

\section*{Acknowledgments}
We thank Steve Warren and Enrique Gazta\~{n}aga 
for helpful discussions.
We also thank the staff of the AAO  for hospitality and efficient observing.
 This work was supported by grants
from the UK Particle Physics and Astronomy Research Council.
RACC acknowledges the receipt of a 
SERC/PPARC studentship and travel grants 
as well as support from NASA Astrophysical Theory Grants
NAG5-2864 and NAG5-3111.

\section*{References}   
\def\refe {\par \hangindent=.7cm \hangafter=1 \noindent}

\refe
Abell, G.~O., Corwin, H.~G.  \& Olowin, R.~P., 1989,
 {\it Ap. J. Suppl.}, {\bf 70}, 1.

\refe
Abell, G.~O., 1958,
 {\it Ap. J. Suppl.}, {\bf 3}, 211.

\refe
Bahcall, N.~A. \& Cen, R., 1992,
 {\it Ap. J. Lett.}, {\bf 398}, L81.

\refe
Bahcall, N.~A. \& Soneira, R.~M., 1983,
 {\it Ap. J.}, {\bf 270}, 20.

\refe
Bahcall, N.~A. \& West, M., 1992,
 {\it Ap. J.}, {\bf 392}, 419.

\refe
Bogart, R.~S. \& Wagoner, R.~V., 1973,
 {\it Ap. J.}, {\bf 181}, 609.

\refe
Croft, R. A.~C. \& Efstathiou, G., 1994,
 {\it Mon. Not. R. astr. Soc.}, {\bf 267}, 390.

\refe
Croft, R. A.~C. \& Efstathiou, G., 1995,
 In: {\it Proceedings of the 11th Potsdam Cosmology Workshop}, ed.
  Muecket, J., World Scientific.

\refe
da Costa, L.~N., Geller, M.~J., Pellegrini, P.~S., Latham, D.~W., Fairall,
  A.~P., Marzke, R.~O., Willmer, C. N.~A., Huchra, J.~P., Calderon, J.~H.,
  Ramella, M.  \& Kurtz, M.~J., 1994,
 {\it Ap. J. Lett.}, {\bf 424}, L1.

\refe
Dalton, G.~B., Efstathiou, G., Maddox, S.~J.  \& Sutherland, W.~J., 1992,
 {\it Ap. J. Lett.}, {\bf 390}, L1 ({\bf DEMS92}).

\refe
Dalton, G.~B., Croft, R. A.~C., Efstathiou, G., Sutherland, W.~J., Maddox,
  S.~J.  \& Davis, M., 1994a,
 {\it Mon. Not. R. astr. Soc.}, {\bf 271}, L47.

\refe
Dalton, G.~B., Efstathiou, G., Maddox, S.~J.  \& Sutherland, W.~J., 1994b,
 {\it Mon. Not. R. astr. Soc.}, {\bf 269}, 151.

\refe
Dalton, G.~B., Efstathiou, G., Sutherland, W.~J., Maddox, S.~J.  \& Davis, M.,
  1997,
 {\it Mon. Not. R. astr. Soc.}, {\bf }, {\it submitted}.

\refe
Davis, M. \& Peebles, P. J.~E., 1983,
 {\it Ap. J.}, {\bf 267}, 465.

\refe
Dekel, A., Blumenthal, G.~R., Primack, J.~R.  \& Olivier, S., 1989,
 {\it Ap. J. Lett.}, {\bf 338}, L5.

\refe
Efstathiou, G., Dalton, G.~B., Sutherland, W.~J.  \& Maddox, S.~J., 1992,
 {\it Mon. Not. R. astr. Soc.}, {\bf 257}, 125 .

\refe
Efstathiou, G., Sutherland, W.~J.  \& Maddox, S.~J., 1990,
 {\it Nature}, {\bf 348}, 705.

\refe
Eke, V., Cole, S., Frenk, C.~S.  \& Navarro, J., 1996,
 {\it Mon. Not. R. astr. Soc.}, {\bf 281}, 703.

\refe
Hamilton, A. J.~S., 1993,
 {\it Ap. J. Lett.}, {\bf 406}, L47.

\refe
Hauser, M.~G. \& Peebles, P. J.~E., 1973,
 {\it Ap. J.}, {\bf 185}, 757.

\refe
Huchra, J.~P., Henry, P., Postman, M.  \& Geller, M.~J., 1990,
 {\it Ap. J.}, {\bf 365}, 66.

\refe
Klypin, A.~A. \& Kopylov, A.~I., 1983,
 {\it Sov. Astron. Lett.}, {\bf 9}, 41.

\refe
Klypin, A., Holtzman, J., Primack, J.  \& Regos, E., 1993,
 {\it Ap. J.}, {\bf 416}, 1.

\refe
Lumsden, S.~L., Nichol, R.~C., Collins, C.~A.  \& Guzzo, L., 1992,
 {\it Mon. Not. R. astr. Soc.}, {\bf 258}, 1.

\refe
Mann, R.~G., Heavens, A.~F.  \& Peacock, J.~A., 1993,
 {\it Mon. Not. R. astr. Soc.}, {\bf 263}, 798.

\refe
Marshall, H.~L., Avni, Y., Tananbaum, H.  \& Zamorani, G., 1983,
 {\it Ap. J.}, {\bf 269}, 35.

\refe
Mo, H.~J., Jing, Y.~P.  \& White, S. D.~M., 1996,
 {\it Mon. Not. R. astr. Soc.}, {\bf }, {\it submitted}.

\refe
Nichol, R.~C., Briel, U.~G.  \& Henry, J.~P., 1994,
 {\it Mon. Not. R. astr. Soc.}, {\bf 267}, 771.

\refe
Nichol, R.~C., Collins, C.~A., Guzzo, L.  \& Lumsden, S.~L., 1992,
 {\it Mon. Not. R. astr. Soc.}, {\bf 255}, 21P.

\refe
Peacock, J.~A. \& West, M.~J., 1992,
 {\it Mon. Not. R. astr. Soc.}, {\bf 259}, 494.

\refe
Peebles, P. J.~E., 1980,
 {\it The Large Scale Structure of the Universe}, Princeton University
  Press.

\refe
Postman, M., Huchra, J.~P.  \& Geller, M.~J., 1992,
 {\it Ap. J.}, {\bf 384}, 404 ({\bf PHG}).

\refe
Romer, A.~K., Collins, C., MacGillivray, H., Cruddace, R.~G., Ebeling, H.  \&
  H.Boringer, 1994,
 {\it Nature}, {\bf 372}, 75.

\refe
Sutherland, W.~J. \& Efstathiou, G., 1991,
 {\it Mon. Not. R. astr. Soc.}, {\bf 248}, 159.

\refe
Sutherland, W.~J., 1988,
 {\it Mon. Not. R. astr. Soc.}, {\bf 234}, 159.

\refe
Tonry, J. \& Davis, M., 1979,
 {\it Ap. J.}, {\bf 84}, 1511.

\refe
Wright, E.~L., Smoot, G.~F., Kogut, A., Hinshaw, G., Tenorio, L., Lineweaver,
  C., Bennett, C.~L.  \& Lubin, P.~M., 1994,
 {\it Ap. J.}, {\bf 420}, 1.

\end{document}